\documentclass[letters,fleqn,usenatbib]{mnras}

\usepackage{newtxtext,newtxmath}

\usepackage[T1]{fontenc}
\usepackage{ae,aecompl}

\usepackage{graphicx}	
\usepackage{amsmath}	
\usepackage{amssymb}	

\title[Rotation-metallicity gradients in the MW disks]{Evidence of a large scale positive rotation-metallicity correlation in the Galactic thick disk}

\author[P. Re Fiorentin et al.]{
Paola Re Fiorentin,$^{1}$\thanks{E-mail: paola.refiorentin@inaf.it}
Mario G. Lattanzi,$^{1}$
and Alessandro Spagna$^{1}$
\\
$^{1}$INAF - Osservatorio Astrofisico di Torino, Strada Osservatorio 20, 10025 Pino Torinese, TO, Italy
}

\date{Accepted 2018 October 17. Received 2018 October 17; in original form 2018 August 6}

\pubyear{2018}

\begin{document}

\label{firstpage}
\pagerange{\pageref{firstpage}--\pageref{lastpage}}
\maketitle

\begin{abstract}
This study is based on high quality astrometric and spectroscopic data from the most recent releases by 
Gaia and APOGEE. 
We select $58\,882$ thin and thick disk red giants, in the Galactocentric (cylindrical) distance range $5 < R < 13$~kpc and within $|z| < 3$~kpc,    
for which full chemo-kinematical information is available. 
Radial chemical gradients, 
$\partial \rm{[M/H]} / \partial \rm{R}$,
and rotational velocity-metallicity correlations, 
$\partial V_\phi / \partial \rm{[M/H]}$,
are re-derived firmly uncovering that the thick disk velocity-metallicity correlation maintains its positiveness over the $8$~kpc range explored.    
This observational result is important as it sets experimental constraints on recent theoretical studies on 
the formation and evolution of the Milky Way disk and on cosmological models of Galaxy formation.
\end{abstract}

\begin{keywords}
Galaxy: formation --- Galaxy: disk --- Galaxy: abundances --- Galaxy: kinematics and dynamics
\end{keywords}



\section{Introduction}\label{sec:1}

Although the existence of a thick disk in the Milky Way was revealed $35$ years ago \citep{Gilmore1983}
and its spatial, kinematic, and chemical properties are today better defined, its origin is still matter of debate.

Proposed scenarios include the heating of a pre-existing thin disk through a minor merger \citep[][]{Villalobos2008},
accretion of dwarf galaxies stars from disrupted satellites \citep[][]{Abadi2003},
or stars formed in situ from gas-rich mergers at high redshift \citep[][]{Brook2005, Haywood2015}.
On the other hand, simulations suggest that thick disks could be produced
through secular radial migration of stars induced by 
spiral arms \citep[][]{Roskar2008, Schoenrich2009, Curir2012} and 
flaring combined with inside-out disk formation \citep{Minchev2017}.
Such models predict characteristic trends on the kinematics and chemical abundances that can be used
to discriminate the one, or the ones, favoured by the Milky Way.

Indeed, detailed information can be obtained from massive
astrometric and spectroscopic surveys.
Quite recently, the European Space Agency's Gaia mission has made its second data release, or Gaia DR2
\citep[][]{Gaia Collaboration2018}: it provides unprecedented accurate
measurements of parallax and proper motion  
for more than 1.3 billion stars across the whole sky \citep[][]{Lindegren2018}.
On the other hand, as part of SDSS-IV,
the Apache Point Observatory Galactic Evolution Experiment fourteenth data release, APOGEE~DR14, 
has delivered high-resolution ($R \sim 22\,500$)
high signal-to-noise near-infrared spectra,
enabling the determination of precise 
radial velocities as well as stellar parameters
and abundances for more than $20$ chemical elements \citep[][]{Majewski2017, Abolfathi2018}. 
Finally, we recall that the APOGEE Survey targeted mostly red giant stars \citep[][]{Zasowski2013}.

In this Letter we take advantage of these new superb measurements to improve on the recent work of 
\citet[][]{Kordopatis2017} based on APOGEE~DR12. 
Section~\ref{sec:2} describes our selection of disk stars, Section~\ref{sec:3}
presents the findings that resulted from our analysis, 
while Section~\ref{sec:4} addresses the theoretical implications of our results.

\section{Data and sample selection}\label{sec:2}

This study starts with a new kinematic catalogue, 
assembled after cross-matching Gaia~DR2 and APOGEE~DR14.
The resulting sample contains DR2 positions, parallaxes and proper motions \citep[][]{Lindegren2018}, 
plus radial velocities and chemical abundances derived
with the APOGEE Stellar Spectra Parameter Pipeline \citep[e. g.,][]{Holtzman2015,Garcia2016}.
The data set allows us to derive a complete, 6D, phase-space information for a 
sufficiently {\it pure} sample of tracers of the disk (thin and thick) populations.

We first select only objects having astrometric solutions 
that are either not affected by excess of noise, $\epsilon=0$, or with a significance level on $\epsilon$ less than 
two, to discard astrometric binaries and other anomalous cases  \citep[see][for details]{Lindegren2018}.
Then, we retain only those stars
with relative parallax error $\sigma_\varpi/\varpi < 0.2$, that allows to compute
distances as $d = 1/\varpi$ with quasi-gaussian errors.
As for the selection on the APOGEE~DR14 data, 
we reject stars with flags warning of poor stellar parameter 
estimates, and those with signal-to-noise ratios lower
than $70$.
Also, in order to work with reliable $\alpha$-element abundances,
we only consider stars with 
$\chi^2 < 10$ and $4000 <  T_{\rm{eff}}< 5000$~K 
according to \citet[][]{Anders2014}.
Therefore, our initial sample comprises a total of $69\,400$ red giants down to $G=17.73$~mag  
with only $26$ stars fainter than $16.5$ mag.
Median uncertainties are: $0.03~\rm{mas}$ in parallax, 
$50~\rm{\mu as}$ in annual proper motion, 
and $\sim 100~\rm{m~s}^{-1}$ for the APOGEE provided line-of-sight velocities.

Three-dimensional velocities in  Galactocentric cylindrical coordinates, $(V_R, V_\phi, V_z)$, 
are derived by assuming that the Sun is $8.5~\rm{kpc}$ away from the Milky Way (MW) centre, 
the LSR rotates at $232~\rm{km~s}^{-1}$ around the Galactic centre \citep[][]{McMillan2017}, 
and the LSR peculiar velocity of the Sun is $(U,V,W)_{\sun}=(11.1,12.24,7.25)~\rm{km~s}^{-1}$ \citep[][]{Schoenrich2010}. 
Median uncertainties of the derived Galactocentric velocities are 
$(\sigma_{V_{R}},\sigma_{V_{\phi}},\sigma_{V_{z}})=(0.53,0.64,0.54)~\rm{km~s}^{-1}$.

\begin{figure}
   \centering
   \includegraphics[width=\linewidth]{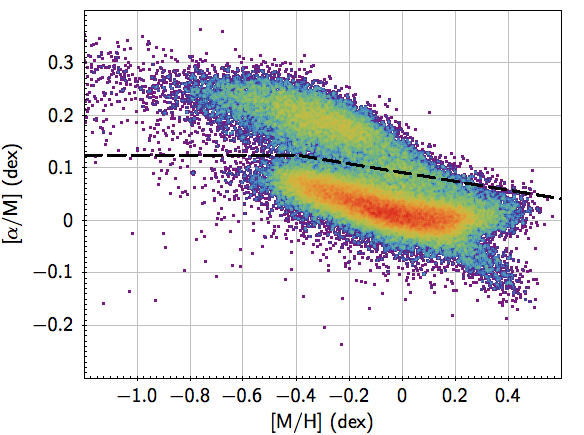}
   \caption{Chemical distribution, $\rm{[\alpha/M]}$  vs. $\rm{[M/H]}$, for the $67\,358$    
   	Gaia~DR2-APOGEE~DR14 stars with $\rm{[M/H]}> -1.2$~dex and $|z| < 3$~kpc.
	The dashed line represents the adopted separation between thin disk (below) and thick disk (above). 
	Typical errors are below $0.03$~dex per $\rm{[\alpha/M]}$ and less than $0.07$~dex per $\rm{[M/H]}$.
              }
   \label{MetallicityAlphaE}
 \end{figure}

Disk stars are chemically selected utilising the constraint 
$\rm{[M/H]} > -1.2$~dex\footnote{We adopt the overall chemical abundance $\rm{[M/H]}$ and 
$\alpha$-element abundances 
$\rm{[\alpha/M]}$, as derived by the APOGEE Stellar Parameter and Chemical Abundances Pipeline \citep[ASPCAP,][]{Zamora2015}. 
The stellar parameters fitted by ASPCAP are $T_{\rm{eff}}$, $\log~g$, $\xi_t$, $\rm{[M/H]}$, $\rm{[\alpha/M]}$,  $\rm{[C/M]}$,  and $\rm{[N/M]}$.  
Here, the overall metallicity is defined as 
$\rm{[M/H]} = \log(N_M/N_H)_{\star} -\log(N_M/N_H)_{\sun}$, 
where $N_M$ and $N_H$ are 
the number density of all elements with atom number $Z > 2$ and hydrogen nuclei, respectively. 
The $\alpha$-elements considered are O, Ne, Mg, Si, S, Ca, and Ti.
}
and $|z| <3$~kpc, resulting in a 
sample of $67\,358$ objects. 
Fig.~\ref{MetallicityAlphaE} shows the chemical plane, $\rm{[\alpha/M]}$ vs. $\rm{[M/H]}$, for this sample;
the thin disk (low-$\alpha$) and thick disk (high-$\alpha$) subsamples 
are reasonably well separated (below and above the dashed line, respectively) by the relation: 

\begin{equation}
\rm{[\alpha/M]}=\left\{
\begin{array}{ll}
+0.125 & \text{if} \,\,\, \rm{[M/H]}< -0.4\\
-0.083 (\rm{[M/H]} -0.5)+0.05 & \text{if} \,\,\, \rm{[M/H]} \ge -0.4
\end{array}\right.
\label{eq:selection}
\end{equation}

\begin{figure}
   \centering
   \includegraphics[width=\linewidth]{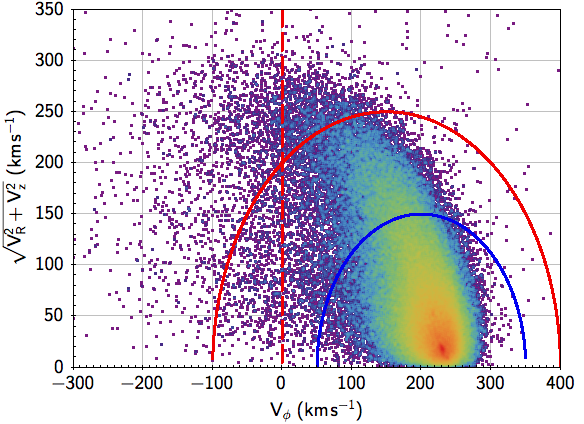}
   \caption{Toomre diagram of the disk sample in Fig.~\ref{MetallicityAlphaE}. 
   	The circles define the Toomre inequalities (see end of Section~\ref{sec:2}) 
	we used to minimise contamination from halo stars for the thin (blue) and the thick (red) disks.
	The vertical line, $V_\phi=0$, is further used to reject halo (retrograde) stars that can contaminate the thick disk sample.
	}
   \label{Toomre}
 \end{figure}

In order to minimise kinematic contamination from the halo, we retain only stars with $V_\phi > 0$ 
and apply the Toomre selections (see Fig.~\ref{Toomre}): 
\begin{eqnarray}
(V_\phi-200)^2+V_R^2+V_z^2 & < & 150^2~\rm{km^2~s^{-2}}\label{eq:2}\\
(V_\phi-150)^2+V_R^2+V_z^2 & < &  250^2~\rm{km^2~s^{-2}}\label{eq:3}
\end{eqnarray}
for thin disk (Eq.~\ref{eq:2}) and thick disk (Eq.~\ref{eq:3}) stars.
With this further selection our sample is composed of 
$60\,539$ disk stars.

Finally, we remind the reader that the photometric selection criteria of the APOGEE targets and the $T_{\rm{eff}}$-$\log~g$ 
boundaries of the ASPCAP grid may bias the resulting metallicity distribution, 
e.g. by undersampling the metal rich tail at $\rm{[M/H]}> +0.1$~dex, as discussed by \citet[][]{Hayden2014}. 
The radial metallicity gradients and the velocity metallicity correlations, especially for the thin disk, could be sensitive to this effect and in particular at larger distances.
We briefly address this issue in the next Section.

\section{Results}\label{sec:3}

We focus on the sample within the 
Galactocentric (cylindrical) distance range $5 < R < 13$~kpc for a total of $58\,882$ disk stars. 
Also, the following analysis refers to $4$ Galactocentric rings, each $2-\rm{kpc}$ wide, 
with central radius at $R=6, 8, 10, 12$~kpc, respectively. 
In addition, both chemically selected thin and thick disk stars are considered 
in three height intervals:  $|z| <3$~kpc, $|z| <1$~kpc, and $1 \le |z| <3$~kpc. 

\begin{figure}
   \centering
   \includegraphics[width=\linewidth]{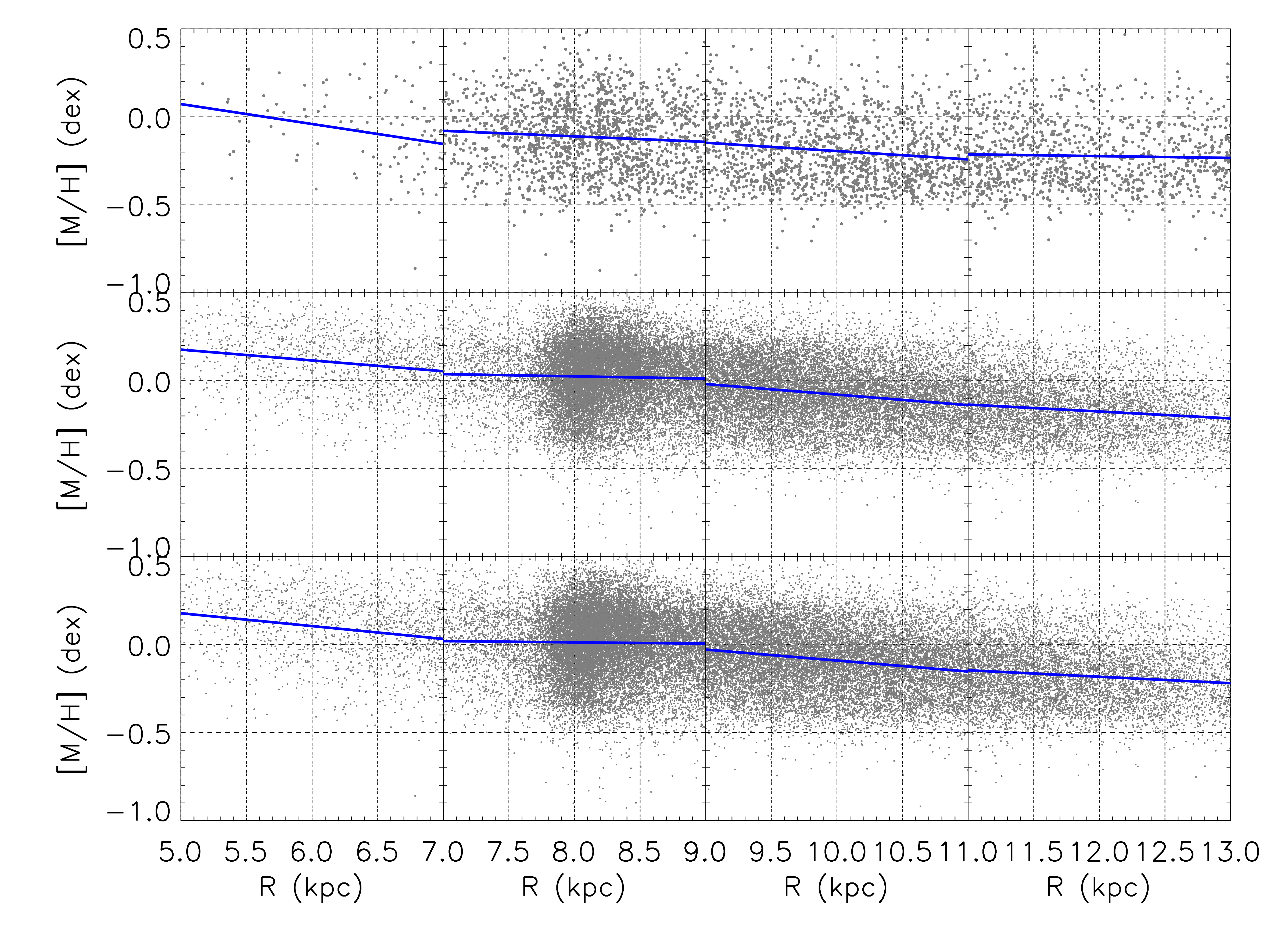}
   \caption{Radial metallicity distribution of thin disk stars  
   	in the Galactocentric (cylindrical) distance range $5 < R < 13$~kpc and within $|z| < 3$~kpc. 
   	From left to right, the full radial range is shown in four $2-\rm{kpc}$ wide rings at $R=6, 8, 10, 12$~kpc.
	The sample is also divided in three height intervals: $|z| <3$~kpc (bottom panels), $|z| < 1$~kpc (middle panels), and $1 \le |z| <3$~kpc (top panels). 
	 The solid lines represent the linear fits to the data.
              }
   \label{MH-R_th}
 \end{figure}
\begin{figure}
   \centering
   \includegraphics[width=\linewidth]{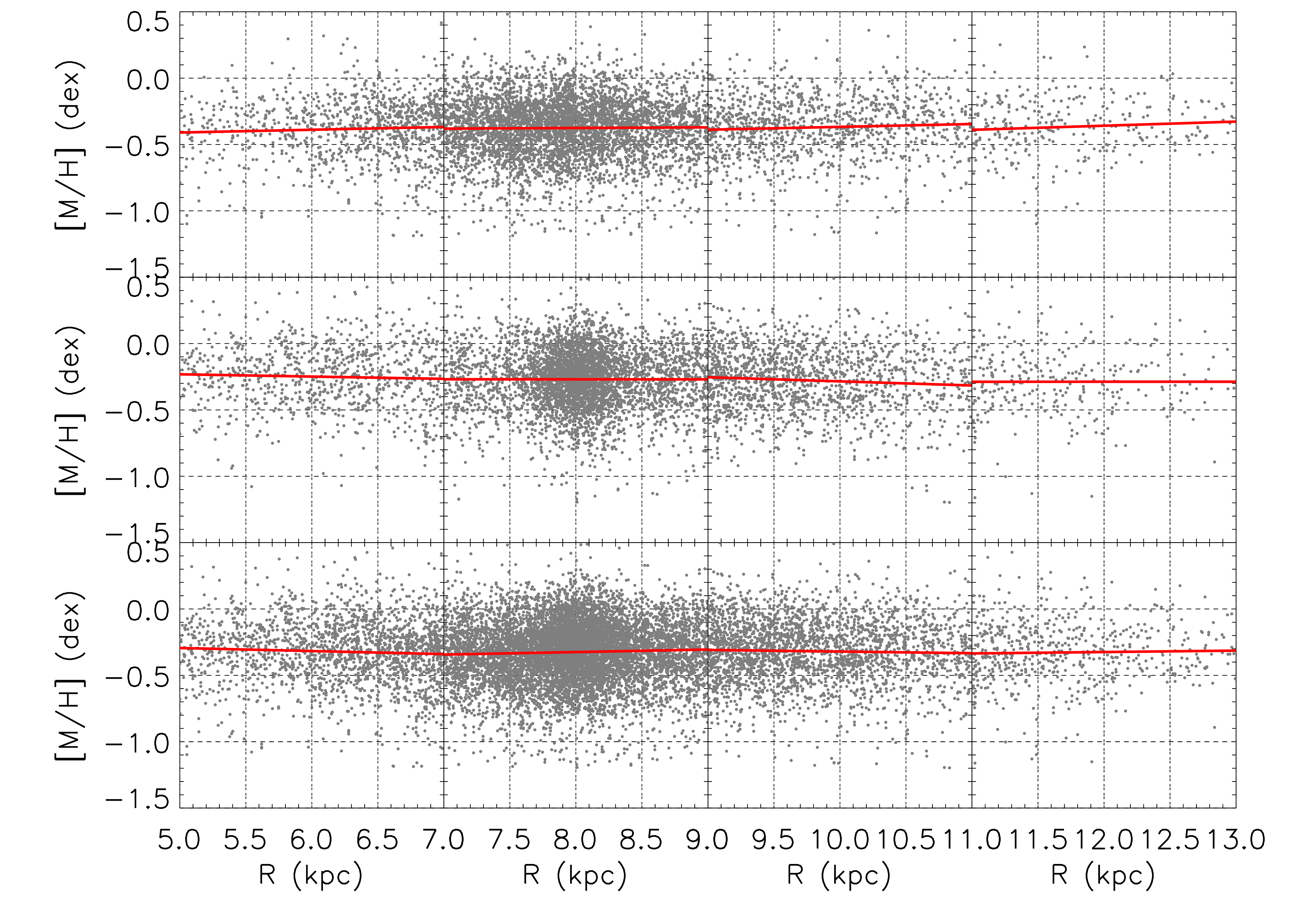}
   \caption{As Fig.~\ref{MH-R_th}, but for thick disk stars.
              }
   \label{MH-R_TK}
 \end{figure}

Figs.~\ref{MH-R_th}--\ref{MH-R_TK} show the distribution of $\rm{[M/H]}$ as a function of $R$, and the three $z$ intervals above,  
for thin and thick disk stars, respectively. 
A linear fit to the data is also shown for each of the Galactocentric rings examined, 
while the actual fit results are provided in Table~\ref{table:1}.
At a given range, radial gradients appear consistent with trends expected for the two disk populations irrespective of distance from the Galactic plane. 
In particular, thick disk stars show rather flat (or mildly positive) gradients throughout the portion of the disk proved.
Within the Solar annulus,  $7 < R < 9$~kpc, we measure for the low-$\alpha$ (thin disk) population a radial-metallicity gradient of  
$-0.031\pm 0.010~\rm{dex~kpc^{-1}}$ between $1\le |z|<3$~kpc, and of $-0.013\pm 0.003~\rm{dex~kpc^{-1}}$ below $|z|<1$~kpc.
For the high-$\alpha$ population, the fit estimations go from a flat slope of $0.000\pm 0.008~\rm{dex~kpc^{-1}}$  for $|z| < 1$~kpc 
to a positive slope of $0.019\pm 0.005~\rm{dex~kpc^{-1}}$  below $|z| = 3$~kpc.
These gradients are consistent with recent measurements like in, e.g., \citet[][]{Anders2014,Recio-Blanco2014,Li2017, Peng2018}.

\begin{figure*}
   \centering
   \includegraphics[width=0.84\linewidth]{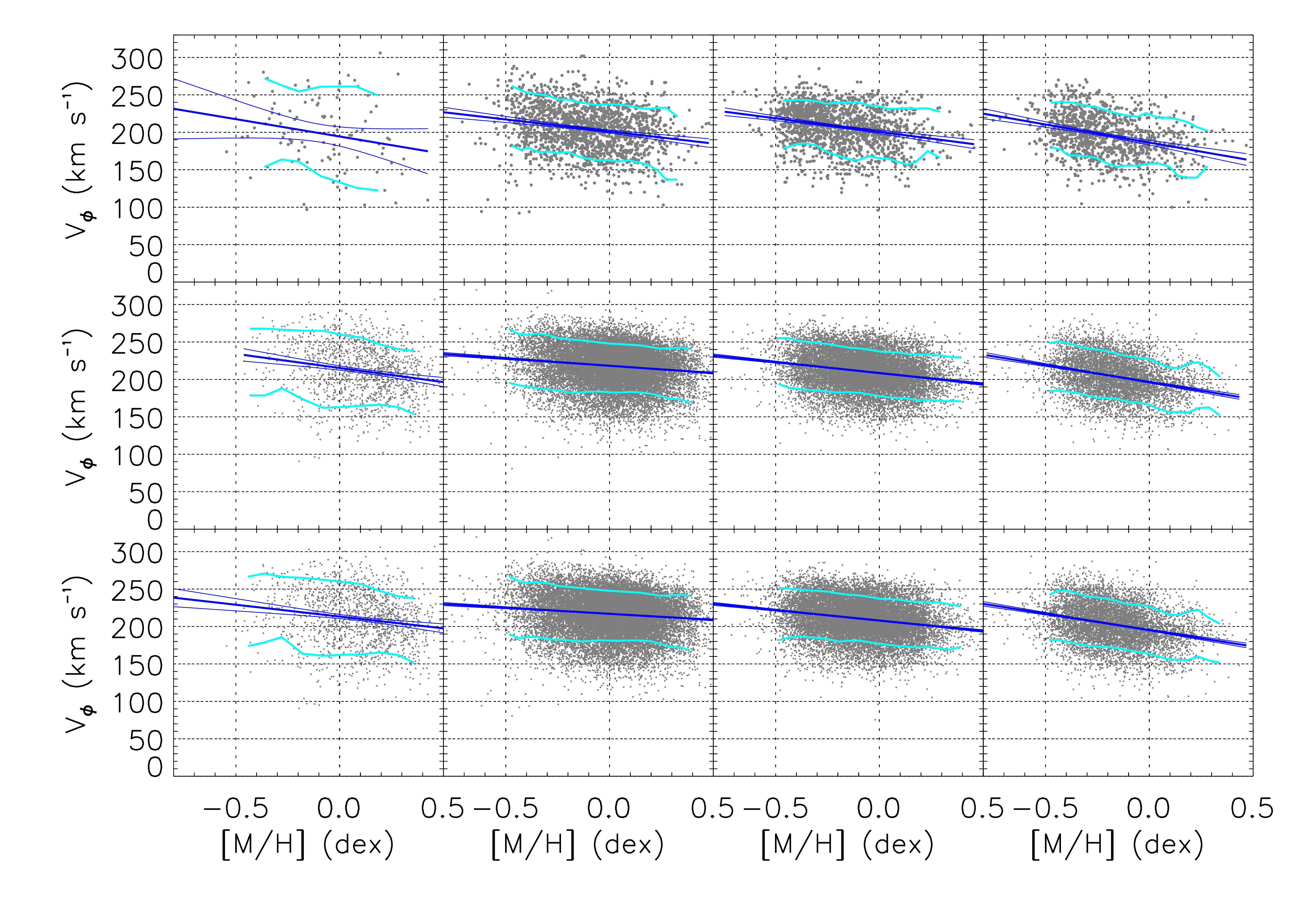}
   \caption{Velocity-metallicity distribution of thin disk stars as function of 
   	Galactocentric (cylindrical) radius $R$, and distance from the plane $|z|$.
	The sample is shown, from left to right, in four $2-$~kpc wide rings at the central radii $R=6, 8, 10, 12$~kpc. 
	In the vertical direction, the bottom panels show stars with $|z| <3$~kpc, the middle panels stars with $|z| <1$~kpc, 
	while the parts of the sample with $1 \le |z| <3$~kpc is given in the top panels.
	The solid blue lines represent the linear fits to the data and the corresponding $99$\% confidence level curves. 
	The solid cyan lines are the $10$th and $90$th percentiles of the data shown.
              }
   \label{Vphi-MH_th}
 \end{figure*}

\begin{figure*}
   \centering
   \includegraphics[width=0.84\linewidth]{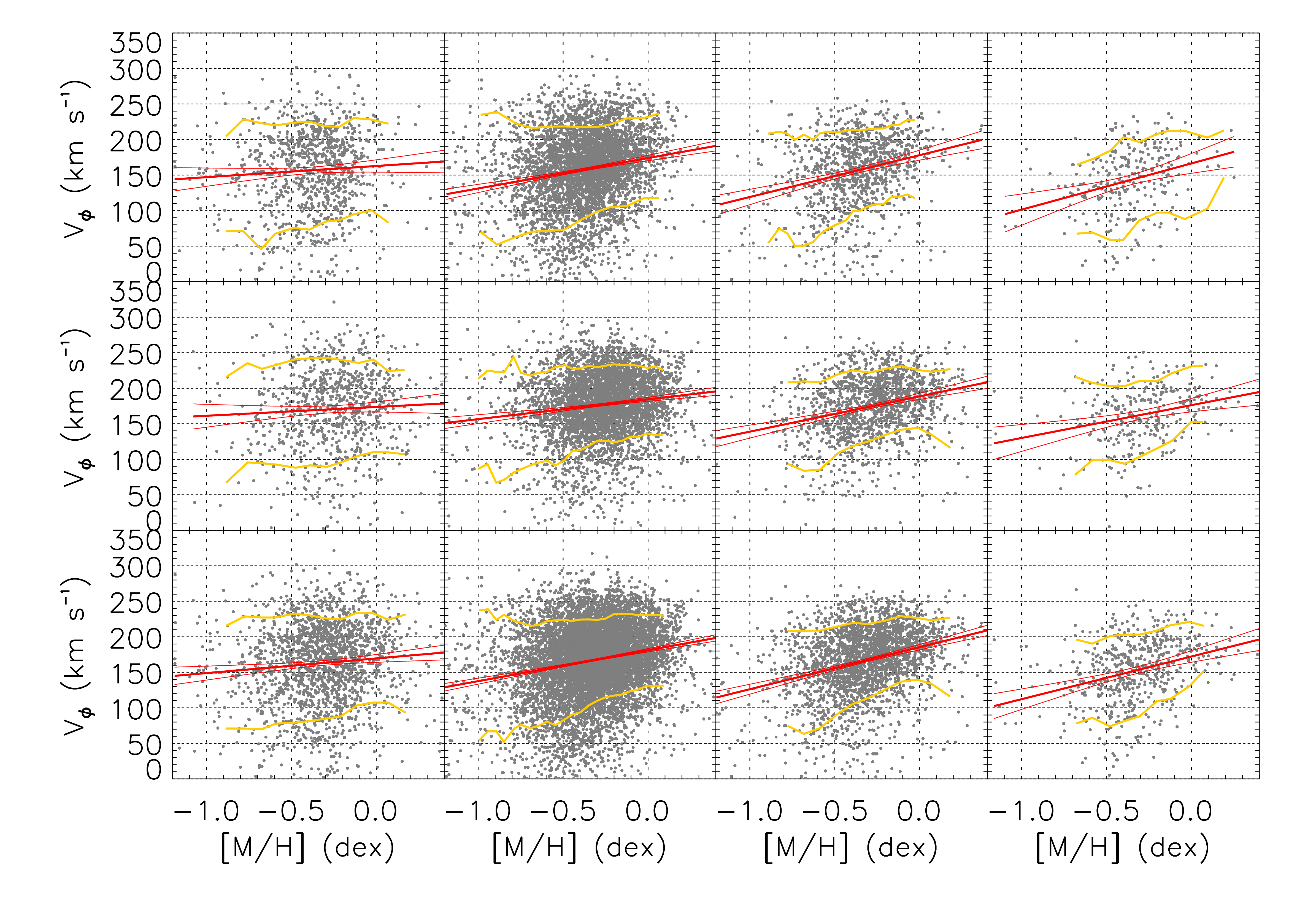}
   \caption{	As Fig.~\ref{Vphi-MH_th}, but for thick disk stars.
              }
   \label{Vphi-MH_TK}
 \end{figure*}

Figs.~\ref{Vphi-MH_th}--\ref{Vphi-MH_TK} show the distribution of rotational velocity $V_\phi$ with $\rm{[M/H]}$ 
for thin and thick disk stars, respectively. 
The slope, $\partial V_\phi / \partial \rm{[M/H]}$, is estimated again 
via a linear fit to the data for each of the four MW rings and the three Galactic plane strips defined above. 
Table~\ref{table:2} provides the results of the fits and lists: total number of stars, 
slope (i.e., the velocity-metallicity gradients), and Spearman's rank correlation coefficient 
for each of the bin (in $R$ and $|z|$) studied.
We clearly notice {\it negative} velocity-metallicity correlations for thin disk stars at any $|z|$ and throughout the disk;
the thick disk population shows quite a similar behaviour but for {\it positive} correlations.
Moreover, as expected, $V_\phi$ slows down with increasing $|z|$ for both chemically selected populations. 

At the Solar circle, i.e. $7 < R < 9$~kpc, 
a kinematical-metallicity correlation of  
$42.8\pm 3.4~\rm{km~s^{-1}~dex^{-1}}$ is estimated for the thick disk between $1\le |z|<3$~kpc,
while a shallower slope $27.9\pm 3.2~\rm{km~s^{-1}~dex^{-1}}$ is present at $|z|<1$~kpc (see Fig.~\ref{Vphi-MH_TK}, second panels from the left). 
These estimates are quite compatible with the 
earlier measurements of \citet[][]{Spagna2010}, and later confirmed by a number of independent studies 
\citep[][]{Lee2011,Kordopatis2011,Adibekyan2013, Haywood2013,Kordopatis2013,Guiglion2015,AllendePrieto2016, Wojno2016}.

Finally, let us briefly address the issue of possible biases   
related to the known under-sampling of the metal rich tail of the APOGEE Survey (see end of Sect.~\ref{sec:2}). 
We tested the robustness of the findings above by repeating all of our linear least-squares fits 
after applying metallicity cuts $\rm{[M/H]< [M/H]_{max}}$, with $0.0~\rm{dex} \le \rm{[M/H]_{max}}\le 0.5~\rm{dex}$, to thin and thick disk star samples alike.
Within the errors, all the fits appear quite consistent with the results we present in Tables~\ref{table:1} and \ref{table:2}; in other words, 
we do not find any statistically significant (propagated) bias affecting either the radial metallicity gradients or the rotational-metallicity correlations of both the thin and thick disks.

\begin{table*}
	\centering
	\caption{Radial metallicity gradients, $\partial \rm{[M/H]} / \partial \rm{R}$ in $\rm{dex~kpc^{-1}}$, in bins of $R$ and $|z|$, for disk stars below $|z|= 3$~kpc.}
	\label{table:1}
	\begin{tabular}{lccccccccc}
		\hline
		Sample & $\rm{|z|~kpc}$ & \multicolumn{2}{c}{$\rm{R=6~kpc}$}& \multicolumn{2}{c}{$\rm{R=8~kpc}$}& \multicolumn{2}{c}{$\rm{R=10~kpc}$}& \multicolumn{2}{c}{$\rm{R=12~kpc}$}\\
			     &          & N & $\partial \rm{[M/H]} / \partial \rm{R}$ & N & $\partial \rm{[M/H]} / \partial \rm{R}$ & N & $\partial \rm{[M/H]} / \partial \rm{R}$ & N & $\partial \rm{[M/H]} / \partial \rm{R}$\\
		\hline
   Thin disk     & $\ge 1$   &    $   99$      &  $-0.113\pm 0.044$   &   $1688$  &   $-0.031\pm 0.010$     &   $1557$  &   $-0.047\pm 0.009$   & $1078$  &   $-0.010\pm 0.010$\\
                      & $<1$   &    $1291$     &  $-0.061\pm 0.009$   & $19585$  &   $-0.013\pm 0.003$     & $13985$  &   $-0.060\pm 0.003$   & $5690$  &   $-0.039\pm 0.004$\\
                      & $\rm{all}$   &    $1390$     &  $-0.073\pm 0.010$   & $21273$  &   $-0.007\pm 0.003$     & $15542$  &   $-0.062\pm 0.003$   & $6768$  &   $-0.037\pm 0.004$\\ 
                      
   Thick disk   & $\ge1$   &    $1048$     &  $+0.021\pm 0.014$  &   $4612$  &   $+0.005\pm 0.007$     &   $1113$  &   $+0.022\pm 0.012$  & $ 284$  &    $+0.031\pm 0.023$\\ 
                      & $<1$   &    $ 863$      &  $-0.017\pm 0.015$   &   $4270$  &   $ 0.000\pm 0.008$      &   $1438$  &   $-0.031\pm 0.011$   & $ 281$  &    $ 0.000\pm 0.030$\\ 
                      & $\rm{all}$   &    $1911$      &  $-0.023\pm 0.010$   &   $8882$  &   $+0.019\pm 0.005$      &   $2551$  &   $-0.012\pm 0.009$   & $565$  &    $+0.011\pm 0.019$\\ 
		\hline
	\end{tabular}
\end{table*}

\begin{table*}
	\centering
	\caption{Rotational-metallicity correlations, $\partial V_\phi / \partial \rm{[M/H]}$ in  $\rm{km~s^{-1}~dex^{-1}}$, in bins of $R$ and $|z|$, for disk stars below $|z|= 3$~kpc.
	}
	\label{table:2}
	\begin{tabular}{lccccccccccccc}
		\hline
		Sample & $\rm{|z|~kpc}$ & \multicolumn{3}{c}{$\rm{R=6~kpc}$}& \multicolumn{3}{c}{$\rm{R=8~kpc}$}& \multicolumn{3}{c}{$\rm{R=10~kpc}$}& \multicolumn{3}{c}{$\rm{R=12~kpc}$}\\
		             &	 &  N & $\partial V_\phi / \partial \rm{[M/H]}$ & $\rho_s$ & N & $\partial V_\phi / \partial \rm{[M/H]}$ & $\rho_s$ & N & $\partial V_\phi / \partial \rm{[M/H]}$ & $\rho_s$ & N & $\partial V_\phi / \partial \rm{[M/H]}$ &  $\rho_s$\\
		\hline
		Thin  &  $\ge 1$   &    $  99$   &  $-46.4\pm 20.5$  & $-0.22$  &  $1688$  &   $-32.2\pm 3.6$   & $-0.23$ &    $1557$ &   $-36.1\pm 3.4$  & $-0.27$  &  $1078$ &   $-48.4\pm 4.3$    & $-0.33$\\
	                          &  $<1$   &    $1291$  &  $-37.7\pm 5.4  $ & $-0.20$  & $19585$ &    $-19.5\pm 1.0$   & $-0.13$ &  $13985$ &   $-29.3\pm 1.1$  & $-0.22$  &  $5690$ &   $-45.8\pm 1.9$    & $-0.30$\\ 
	                          &  $\rm{all}$   &    $1390$  &  $-31.6\pm 5.1  $ & $-0.18$  & $21273$ &    $-16.3\pm 1.0$   & $-0.11$ &  $15542$ &   $-27.9\pm 1.0$  & $-0.22$  &  $6768$ &   $-43.8\pm 1.8$    & $-0.29$\\ 	 
	                          
	        Thick &  $\ge 1$   &    $1048$  &  $+15.7\pm 7.6  $ & $+0.06$  &  $4612$  &   $+42.8\pm 3.4$ & $+0.18$ &   $1113$ &    $+58.9\pm 6.3$ & $+0.27$ &   $ 284$ &   $+65.1\pm 12.7$ & $+0.33$\\ 
	                         &  $<1$   &    $ 863$   &  $+12.4\pm 7.9  $ & $+0.07$  &  $4270$  &   $+27.9\pm 3.2$ & $+0.13$ &   $1438$ &    $+49.7\pm 4.7$ & $+0.27$ &   $ 281$ &   $+46.2\pm  9.7$  & $+0.29$\\
	                         &  $\rm{all}$   &    $1911$   &  $+20.7\pm 5.3  $ & $+0.10$  &  $8882$  &   $+43.5\pm 2.3$ & $+0.20$ &   $2551$ &    $+59.2\pm 3.7$ & $+0.30$ &   $565$ &   $+60.1\pm  7.8$  & $+0.34$\\	                                                   
		\hline
	\end{tabular}
\end{table*}


\section{Discussion and conclusions}\label{sec:4}

By combining astrometric information from Gaia~DR2 with radial velocities and chemical abundances from APOGEE~DR14,
we measured the radial metallicity gradients and, for the first time, 
the rotation metallicity correlation of the MW disk populations as function of $R$ in 
a relatively wide radial range spanning from $5$ to $13$~kpc. 

We adopted the usual chemical classification method of thin and thick disk stars based on 
two main evolutionary sequences in the $\rm{[\alpha/M]}$ vs. $\rm{[M/H]}$ diagram (Fig.~\ref{MetallicityAlphaE}). 
The fact that the chemo-kinematical properties of these two populations are consistent
near the Galactic plane and above $1$~kpc from it
shows the power of the $\alpha$-elements-based classification criterion. 

Concerning the radial metallicity gradient, 
our analysis of the new data confirm the negative slope for the thin disk 
and the flat (or mildly positive) slope for the thick, as found in previous studies based mainly on spectro-photometric distances 
\citep[e.g., ][]{Carrel2012,Anders2014,Recio-Blanco2014,Li2017}.

As for the velocity-metallicity correlations, $\partial V_\phi / \partial \rm{[M/H]}$, that of the thin disk appears
consistently {\it negative} throughout the radial range probed.
This is in line with expectations 
given the negative sign of the corresponding chemical gradient exhibited in Fig.~\ref{MH-R_th}, and 
the radial oscillation of the individual stars caused by the epicyclic component
due to blurring effects \citep[e.g., ][and references therein]{Schoenrich2017,Kawata2018}.  

The situation appears substantially different for the thick disk. 
To the best of our knowledge, Fig.~\ref{Vphi-MH_TK} offers for the first time 
observational evidence that the circular velocity-metallicity correlation of the thick disk is persistently {\it positive} within the 
$8$~kpc range of Galactocentric distances investigated, 
in spite of a quasi-flat metallicity gradient.
This suggests that the chemo-kinematical mechanisms in place for the thin disk are probably replaced, totally or in part, 
by other processes (e.g., inside out formation) when dealing with the $\alpha$-enanced population 
as discussed by \citet[][]{Schoenrich2017} and \citet[][]{Kawata2018}.
Their simulations point out that the present-day radial metallicy and rotation-metallicity correlation 
of the ancient disk stars reflect the imprints of the cosmological conditions (i.e., the original metallicity gradient of the ISM and inside-out formation),
convolved with the mechanisms of secular dynamical evolution of the MW disk, 
and with possible perturbations from satellite mergings.\\ 
To this regard, it is quite interesting to compare our Fig.~\ref{Vphi-MH_TK} to Fig.~$8$ in \citet[][]{Kawata2018}, 
whose first and third rows show a constant positive correlation vs. $R$, which is associated with an almost flat  present-day radial metallicity gradient,     
as per our Fig.~\ref{MH-R_TK}.   
However, if we look at the two figures in greater detail, 
we notice that the observed thick disk rotation-metallicity correlation increases as a function of the galactic radius, 
while the simulations show an opposite trend.\\
\citet[][]{Kawata2018} results derive from N-body MW-like simulations (their C1 and C2-thick1 models) 
that assume positive cosmological chemical gradients in the disk, as initially proposed by \citet[][]{Curir2012}.
Finally, alternative scenarios, as those suggested by the analytical models of \citet[][]{Schoenrich2017}, 
also deserve further investigations possibly employing fully realistic cosmological simulations of the MW as in \citet[][]{Murante2015}. 

There is no doubts that the results presented here will help setting crucial constraints 
on the origin and destiny of the Galactic disk in the context of cosmological models of MW formation.


\section*{Acknowledgements}
We wish to thank the referee for a careful reading of this manuscript and for the useful comments that helped us improve on the original submission. 
This work has made use of data from the European Space Agency (ESA) mission Gaia (https://www.cosmos.esa.int/gaia), 
processed by the Gaia Data Processing and Analysis Consortium (DPAC, https://www.cosmos.esa.int/web/gaia/dpac/consortium). Funding for the DPAC has been provided by national institutions, in particular the institutions participating in the Gaia Multilateral Agreement. 
This work has been funded in part by the Italian Space Agency (ASI) under contract No. 2014-025-R.1.2015 ``Gaia Mission - The Italian Participation to DPAC".




\bsp	
\label{lastpage}
\end{document}